\documentclass[aip,jcp,reprint,twocolumn,superscriptaddress,showpacs,floatfix]{revtex4-2}
\usepackage{mathrsfs}
\usepackage{amssymb}
\usepackage{amsmath}
\usepackage{accents}
\usepackage[nodisplayskipstretch]{setspace}
\usepackage{parskip}
\usepackage{graphicx}  % Include the graphicx package
\usepackage{braket}
\usepackage{comment}
\usepackage{multirow}
\usepackage{url}
\usepackage[colorlinks]{hyperref}
\hypersetup{
    colorlinks=true,    % Enables colored links
    linkcolor=blue,     % Color of internal links
    citecolor=magenta,      % Color of citations
    pdfborder={0 0 0}   % Removes the red boxes around links
}

\begin{document}
\title{Relativistic unitary coupled cluster method for ground-state molecular properties}

\author{Kamal Majee}
\affiliation{\mbox{Department of Chemistry, Indian Institute of Technology Bombay, Powai, Mumbai 400076, India}}
\author{Somesh Chamoli}
\affiliation{\mbox{Department of Chemistry, Indian Institute of Technology Bombay, Powai, Mumbai 400076, India}}
\author{Malaya K. Nayak}
\thanks{Corresponding author}
\email[e-mail: ]{mknayak@barc.gov.in} 
\affiliation{\mbox{Theoretical Chemistry Section, Bhabha Atomic Research Center, Trombay, Mumbai 400085, India}}
\affiliation{\mbox{Homi Bhabha National Institute, BARC Training School Complex, Anushaktinagar, Mumbai 400094, India}}

\author{Achintya Kumar Dutta}
\thanks{Corresponding author}
\email[e-mail: ]{achintya@chem.iitb.ac.in}
\affiliation{\mbox{Department of Chemistry, Indian Institute of Technology Bombay, Powai, Mumbai 400076, India}}
\affiliation{ \small Department of Inorganic Chemistry, Faculty of Natural Sciences, Comenius University, Ilkovičova 6, Mlynská dolina 84215 Bratislava, Slovakia \\
}
\begin{abstract}
\vspace{0.1cm}
We propose a relativistic unitary coupled cluster (UCC) expectation-value approach for computing first-order properties of heavy-element systems. Both perturbative and non-perturbative commutator-based formulations are applied to evaluate ground-state properties, including the permanent dipole moment (PDM), magnetic hyperfine structure (HFS) constant, and electric field gradient (EFG). The results are compared with available experimental data and with those obtained using conventional coupled cluster (CC) calculations. The non-perturbative commutator-based approach truncated at the singles-and-doubles level exhibits markedly better agreement with both the coupled cluster Z-vector method and experimental values than the perturbative variant, likely due to its improved treatment of relaxation effects.
\end{abstract}

\maketitle

\section{\label{sec:level1}Introduction\protect\\}
A comprehensive understanding of chemical processes requires consideration of not only molecular energetics but also intrinsic properties such as dipole moments, hyperfine coupling constants, and other measurable parameters. Therefore, quantum chemical calculations of atomic and molecular properties have become a fundamental aspect of modern computational chemistry.\cite{helgaker2012recent,neese2009prediction,wilson2013methods,norman2018principles}
However, accurate calculation of properties of atoms and molecules containing heavy elements is highly challenging due to the necessity of incorporating both relativistic and electron correlation effects in a balanced manner, especially for systems in which relativistic effects are significant.\cite{visscher1996formulation} One of the most effective ways to account for relativistic effects in a quantum mechanical calculation for many-electron systems is through the Dirac–Hartree–Fock (DHF) method.\cite{dyallIntroductionRelativisticQuantum2007,reiher2014relativistic} The DHF method does not account for the dynamic interaction of opposite-spin electrons, which is known as the electron correlation. Among the various electron correlation methods available, the single-reference coupled cluster (CC) method \cite{cizek1967advances,mukherjee1989use,bartlett1978many,bartlett2007coupled,crawford2007introduction,shavitt2009many} has emerged as one of the most accurate and systematically improvable approaches. The property calculation within the CC method can be performed using two alternative approaches: an expectation value approach,\cite{pal1984variational,bartlett1988expectation} and an analytic derivative technique.\cite{Monkhorst, Salter, Scheiner, Koch90} It should be noted that the two approaches yield different results even for first-order properties, except in the full CC limit.\cite{pal1984variational} 
In the domain of relativistic CC method, the calculation of first-order properties has been reported\cite{visscher1995kramers,visscher1996formulation,petrov2002calculation,eliav1994open,eliav1994ionization,eliav1995relativistic,shee2016analytic,sasmal2015relativistic,prasannaa2016permanent,sasmal2015implementation,sur2005comparative} using both analytic derivative and expectation value approaches. The analytic derivative approach is generally more advantageous for geometric derivatives and higher-order properties.\cite {Koch90} The calculation of second and higher-order properties within the framework of the relativistic CC method has been reported recently.\cite{yuan2024formulation,yuan2023frequency,chakraborty2024spin,4clrccsd} 
In addition to the standard formulation of the CC method,  alternate ansatz exist within the CC framework.\cite{alternatecc} 
One can use a unitary ansatz\cite{Kutzelnigg1977} to arrive at a Hermitian formulation of coupled cluster theory.\cite{NogaUCC,PalUCC84,Hoffmannucc}  In recent years, the unitary coupled cluster (UCC) method has gained significant attention, not only for its computational advantages but also because of its potential applications in quantum computing.\cite{mcardle2020quantum,cao2019quantum,bauer2020quantum,bauer2020quantum} The UCC method has a particular advantage: first-order properties calculated using both expectation-value and analytic-derivative approaches yield the same result for a given truncation of the cluster operator.\cite{PalUCC84} 
Moreover, because the standard CC energy functional is non-Hermitian, the calculation of first-order ground-state properties with the analytic derivative technique requires not only the CC amplitudes but also an additional set of left-vector amplitudes to ensure that the energy functional is stationary. In the UCC expectation value formalism, first-order properties can be obtained using a single set of amplitudes. 
However, unlike conventional CC theory, the UCC framework does not allow a natural truncation of the similarity-transformed Hamiltonian. As a result, an explicit truncation is required. Any arbitrary truncation of the UCC  functional may compromise the size extensivity of the energy.\cite{PalUCC84} To address this challenge, Bartlett and co-workers\cite{bartlett1989alternative} proposed a truncation scheme based on perturbative analysis of the UCC energy functional, thereby preserving size extensivity. Building on this work, Taube and Bartlett introduced an improved truncation scheme that yields exact results for two-electron systems.\cite{taube2006new}  However, for larger and more complex molecules, the perturbative approximation to the UCC theory [UCC(n)] often fails due to the poor convergence behavior of the underlying low-order M\o ller-Plesset(MP) series.\cite{phillips2025comparing} To overcome this limitation, recent efforts have focused on truncating the UCC expansion based on the rank of nested commutators\cite{sur2008relativistic,liu2021unitary,liu2018unitary} using the Bernoulli expansion, which has shown improved performance in systems with irregular MP convergence. Furthermore, Liu \textit{et al.}\cite{liu2018unitary} proposed the UCCn method, which applies a perturbative truncation of the Bernoulli expansion and offers an alternate hierarchy of perturbative approximations to UCC  methods. The UCC3 method has been particularly well studied\cite{liu2018unitary, UCC3EE,UCCprop,UCC3IP} in recent years for ground and excited states of small molecules. Recent work by DePrince and co-workers\cite{phillips2025comparing} demonstrated that the commutator-based truncation scheme achieves faster convergence toward the full configuration interaction (FCI) limit and provides more reliable results than perturbative truncation-based approximations, especially for molecular systems away from their equilibrium geometries. Despite its many attractive features, reports of property calculations using UCC remain limited. Bartlett and co-workers have reported an analytic gradient technique for the UCC(4) method.\cite{WATTS1989359} Sur \textit{et al.} have used the relativistic UCC3 method to calculate nuclear quadrupole moments, hyperfine constants, and transition properties of atoms.\cite{sur2008relativistic} Dreuw and co-workers\cite{UCCprop} have recently reported ground- and excited-state properties using the UCC3 method. One of the main issues with property calculations using low-order Møller–Plesset (MP)-based UCC methods is that they yield less accurate results than the standard CC approach.\cite{WATTS1989359} The non-perturbative quadratic unitary coupled cluster (qUCC) developed by Cheng and co-workers has been shown to provide performance comparable to that of the standard CC method for energy calculations.\cite{liu2021unitary} We have recently reported a relativistic variant of the qUCC method.\cite{majee2024reduced} This paper aims to perform first-order property calculations using the expectation-value approach. The structure of this paper is as follows: Section II presents the theoretical framework, Section III provides the computational details, and Section IV presents the results of molecular property calculations for selected systems. Finally, conclusions are summarized in Section V.
\section{Theoretical framework}
\subsection{\label{app:subsec}Relativistic Unitary Coupled Cluster Theory}
The DHF method is generally the starting point for all relativistic single-reference electron correlation calculations. It extends the Hartree-Fock method by incorporating the special theory of relativity via the Dirac-Coulomb (DC) Hamiltonian$(\hat{H}_{\text{DC}})$ \cite{reiher2014relativistic}  and for a molecular system, under the Born-Oppenheimer approximation, can be defined as
\begin{equation}
\hat{H}_{\text{DC}} = \sum_{i}^{N}\left[ c {\vec{\alpha}}_i \cdot \vec{p}_i + \beta_i m_0 c^2 + \sum_{A}^{N_{\text{nuc}}} V_{iA} \right]+\sum_{i < j} \frac{1}{r_{ij}} I_4 
\label{eq:1}
\end{equation}
where $\vec{p}_i$ denotes the momentum operator, $m_{0}$ is the rest mass, and $c$ is the speed of light. The operator $V_{ia}$ represents the potential energy of the $i$-th electron in the field of the nucleus \textit{A}. The symbols $\vec{\alpha}$ and $\beta$ are the Dirac matrices, and $I_{4}$ denotes the $ 4 \times 4$ identity matrix. The DHF equations  can be represented in matrix form as
\begin{eqnarray}
\resizebox{0.43\textwidth}{!}{$
\begin{bmatrix}
\hat{V}+\hat{J}-\hat{K} & c(\sigma_{\mathrm{psm}}\cdot\vec{{\textit{p}}})-\hat{K} \\
c(\sigma_{\mathrm{psm}}\cdot\vec{{\textit{p}}})-\hat{K} & \hat{V}-2m_{0}c^{2}+\hat{J}-\hat{K}
\end{bmatrix}
\begin{bmatrix}
\Phi^{L} \\ \Phi^{S}
\end{bmatrix}
= E
\begin{bmatrix}
\Phi^{L} \\ \Phi^{S}
\end{bmatrix}
$}
\label{eq:2}
\end{eqnarray}
In this representation,$\Phi^{L}$ and $\Phi^{S}$ correspond to the large and small components of the four-component (4c) wavefunction, each of the components taking the form of a two-spinor. The operator $\hat{V}$ represents the nuclear-electron interaction. The operators $\hat{J}$ and $\hat{K}$ represent the Coulomb and exchange operators, respectively. The Pauli spin matrices are denoted by $\sigma_{\mathrm{psm}}$. \\
Equation (\ref{eq:1}) can be rewritten in the occupation number representation as 
\begin{equation}
\label{eq:3}
\hat{H}_{\text{DC}}=\sum_{pq}{h_{pq}^{\text{4c}}}\hat{a}_{p}^{\dagger}\hat{a}_{q}+\frac{1}{4}\sum_{pqrs}{g_{pqrs}^{\text{4c}}}\hat{a}_{p}^{\dagger}\hat{a}_{q}^{\dagger}\hat{a}_{s}\hat{a}_{r}=\hat{F}+\hat{V}
\end{equation}
Where $\hat{F}$ is the Fock operator, defined as 
\begin{equation}
\label{eq:4}
\hat{F} = \sum_{pq} h^{4c}_{pq}\,\hat{a}^{\dagger}_{p}\hat{a}_{q} 
+ \sum_{i,pq} g_{piqi}^{\text{4c}}\hat{a}^{\dagger}_{p}\hat{a}_{q} 
\end{equation}
and $\hat{V}$ is the so-called fluctuation potential,
\begin{equation}
\label{eq:5}
\hat{V} = - \sum_{i,pq} g_{piqi}^{\text{4c}}\hat{a}^{\dagger}_{p}\hat{a}_{q} +\frac{1}{4}\sum_{pqrs}{g_{pqrs}^{\text{4c}}}\hat{a}_{p}^{\dagger}\hat{a}_{q}^{\dagger}\hat{a}_{s}\hat{a}_{r}
\end{equation}
Here, the indices \textit{p}, \textit{q}, \textit{r}, and \textit{s} represent positive-energy four-component spinors within the framework of the no-pair approximation \cite{PhysRevA.22.348}.
To account for electron-correlation effects, we employ the UCC ansatz on top of the DHF reference state $(\ket{\Phi_0})$.
In the UCC method, the ground-state wave function $\ket{\Psi_{0}}$ is expressed as
\begin{eqnarray}
\ket{\Psi_0}=e^{({\hat{T}-\hat{T}^{\dagger}})}\ket{\Phi_0}
\label{eq:6}.
\end{eqnarray}
The $\hat{T}$ is the excitation operator, while its Hermitian conjugate $\hat{T}^{\dagger}$, accounts for the corresponding de-excitation cluster operator. The difference $\hat{\sigma} = \hat{T}-\hat{T}^{\dagger}$ forms an anti-Hermitian operator, i.e, $\hat{\sigma}=-\hat{\sigma}^{\dagger}$, ensuring that the exponential operator $e^{\hat\sigma}$ is unitary and thus preserves the norm of the wave function.
Within the singles and doubles truncation, the cluster operator is given by 
\begin{equation}
\hat{\sigma} = \hat{\sigma_1}+\hat{\sigma_2}
\label{eq:7}
\end{equation}
\begin{equation}
\hat{\sigma}_1 = \sum_{ia} \left[ {\sigma}_i^a \, \hat{c}_a^\dagger \hat{c}_i - ({\sigma}_i^a)^{*} \, \hat{c}_i^\dagger \hat{c}_a \right]
\label{eq:8}
\end{equation}
\begin{equation}
\hat{\sigma}_2 = \frac{1}{4} \sum_{ijab} \left[ {\sigma}_{ij}^{ab} \, \hat{c}_a^\dagger \hat{c}_b^\dagger \hat{c}_j \hat{c}_i - ({\sigma}_{ij}^{ab})^{*} \, \hat{c}_i^\dagger \hat{c}_j^\dagger \hat{c}_b \hat{c}_a \right]
\label{eq:9}
\end{equation}

where $i$, $j$, $k$, $l$, and $a$, $b$, $c$, $d$ denote the occupied and virtual spinors, respectively.
The ground-state energy is obtained  as 
\begin{equation}
\langle \Phi_0 |\bar {H}| \Phi_0 \rangle = E_{0}
\label{eq:10}
\end{equation}
The cluster amplitudes in the projection based UCC are determined by projecting the electronic Schrödinger equation $\bar{H}\ket{\Phi_{0}}=E_{0}\ket{\Phi_{0}}$ onto the excited state determinants $\bra{\Phi{\mu}}=\bra{\Phi_{i}^{a}},\bra{\Phi_{ij}^{ab}}$ as,
\begin{equation}
\langle \Phi_\mu |\bar {H}| \Phi_0 \rangle =0
\label{eq:11}
\end{equation}
Here, $\bar{H}$ is the similarity-transformed Hamiltonian, defined as
\begin{equation}
\bar {H} = e^{-\hat{\sigma}}\hat{H}_{\text{DC}}e^{\hat{\sigma}}
\label{eq:12}
\end{equation}
This similarity transformed Hamiltonian $(\bar{H})$ can be expanded using the Baker-Campbell-Hausdorff (BCH) expansion formula as
\begin{align}
\bar{H} &= \hat{H}_{\text{DC}} + [\hat{H}_{\text{DC}}, \hat{\sigma}] \nonumber \\
&\quad + \frac{1}{2!} [[\hat{H}_{\text{DC}}, \hat{\sigma}], \hat{\sigma}] \nonumber \\
&\quad + \frac{1}{3!} [[[\hat{H}_{\text{DC}}, \hat{\sigma}], \hat{\sigma}], \hat{\sigma}] + \dots
\label{eq:13}
\end{align}
One major challenge in the UCC theory is that, unlike traditional CC theory, there is no natural finite truncation of the BCH expansion for $\bar{H}$. This arises due to the presence of both excitation and de-excitation operators in $\hat{\sigma}$. Therefore, the expansion requires an explicit truncation.  Special care must be taken to ensure the size extensivity of the truncated unitary coupled cluster method.  This can be achieved either by applying a perturbative truncation\cite{bartlett1989alternative}  based on Møller-Plesset (MP) perturbation theory or by using a commutator-based approach\cite{liu2018unitary}. The commutator-based truncation schemes employing Bernoulli numbers \cite{kut1,kutzelnigg1982quantum} have been found to be particularly effective. These schemes are generally derived using the superoperator (denoted by a double hat) formalism.\cite{prasad1980alternative}  A superoperator ($\hat{\hat{B}}$), when applied to an arbitrary operator ($\hat{A}$), gives rise to a commutator  
\begin{equation}
\hat{\hat{B}} \hat{A}=\left[ \hat{A},\hat{B} \right]
\label{eq:14}
\end{equation}
Using the above definition, Eq. (\ref{eq:12}) can be written as
\begin{equation}
\bar{H}=e^{\hat{\hat{\sigma}}}\hat{H}_{\text{DC}}
\label{eq:15}
\end{equation}
The above expression is equivalent to the BCH expansion given in Eq. (\ref{eq:13}). 
\begin{align}
e^{\hat{\hat{\sigma}}}\hat{H}_{\text{DC}}=\sum_{k=0}^{\infty} \frac{1}{k!} \,\hat{\hat{\sigma}}^k \hat{H}_{\text{DC}}
= \hat{H}_{\text{DC}} + \hat{\hat{\sigma}}\hat{H}_{\text{DC}} + \frac{1}{2}\hat{\hat{\sigma}}^2 \hat{H}_{\text{DC}} + \cdots \nonumber \\
= \hat{H}_{\text{DC}} + [\hat{H}_{\text{DC}}, \hat{\sigma}] + \tfrac{1}{2} [[\hat{H}_{\text{DC}}, \hat{\sigma}], \hat{\sigma}] + \cdots
\label{eq:16}
\end{align}
The operator $\bar{H}$ can now be separated using the superoperator formalism as
\begin{equation}
\bar{H}=e^{\hat{\hat{\sigma}}}\hat{F}+e^{\hat{\hat{\sigma}}}\hat{V}=F+\hat{X}(\hat{\hat{\sigma}})\hat{\hat{\sigma}}\hat{F}+e^{\hat{\hat{\sigma}}}\hat{V}
\label{eq:17}
\end{equation}

Here, $\hat{X}$ denotes the exponential Taylor series expressed as
\begin{equation}
X(\hat{\hat{\sigma}}) = 1 + \frac{1}{2}\hat{\hat{\sigma}} + \frac{1}{6}\hat{\hat{\sigma}}^{2} + \frac{1}{24}\hat{\hat{\sigma}}^{3}+\frac{1}{120}\hat{\hat{\sigma}}^{4}+\cdots
\label{eq:18}
\end{equation}

The inverse of this function can be expressed as
\begin{equation}
X^{-1}(\hat{\hat{\sigma}}) = 1 + \sum_{n>0} B_{n}\,\hat{\hat{\sigma}}^{n}
\label{eq:19}
\end{equation}
Here, $B_{n}$ denotes the Bernoulli numbers 
\begin{equation}
\begin{aligned}
&\quad B_{1} = -\tfrac{1}{2}, \\
&\quad B_{2} = \tfrac{1}{12}, \\
& \quad B_{3} = 0, \\
&\quad  B_{4} = -\tfrac{1}{720}, \\
&\quad \; \ldots
\label{eq:20}
\end{aligned}
\end{equation}
By left-multiplying Eq. (\ref{eq:17}) by $X^{-1}(\hat{\hat{\sigma}})$, one obtains
\begin{equation}
X^{-1}(\hat{\hat{\sigma}})\,[\bar{H} - \hat{F}]
= \hat{\hat{\sigma}} \hat{F} + X^{-1}(\hat{\hat{\sigma}})\, e^{\hat{\hat{\sigma}}} \hat{V}
\label{eq:21}
\end{equation}
Expressions for the iterative generation of $\bar{H}$ is
\begin{equation}
\bar{H} = \hat{F} + \bar{V}
\label{eq:22}
\end{equation}
and
\begin{equation}
\bar{V} = \hat{\hat{\sigma}} \hat{F} 
+ X^{-1}(\hat{\hat{\sigma}})\, e^{\hat{\hat{\sigma}}} \hat{V} 
- \sum_{n>0} B_{n}\, \hat{\hat{\sigma}}^{n} \bar{V}
\label{eq:23}
\end{equation}
It can be seen that the iterative generation of  $\bar{H}$  includes only a singly nested commutator in $\hat{F}$, which makes it advantageous compared to other UCC truncation schemes.   To apply the iterative procedure for generating  $\bar{H}$, the operator $\hat{H}_{\text{DC}}$ is partitioned into two parts:
\begin{equation}
\hat{H}_{\text{DC}}=\hat{H}_{N}+\hat{H}_{R}
\label{eq:24}
\end{equation}
Here, ``\textit{N}" denotes the non-diagonal part of the operator, which contains all pure excitation and de-excitation operators, while ``\textit{R}" represents the remaining part obtained after excluding the non-diagonal part. With this definition, the UCC amplitude equations can be rewritten as
\begin{equation}
\langle \Phi_\mu |\bar {H}| \Phi_0 \rangle =\langle \Phi_\mu |\bar {V}_{N}| \Phi_0 \rangle =0
\label{eq:25}
\end{equation}
The Fock operator is assumed to be block diagonal. The iterative formula for the similarity-transformed fluctuation potential can now be written as 
\begin{equation}
\bar{V}^{(k+1)} 
= \hat{\hat{\sigma}} \hat{F} 
+ X^{-1}(\hat{\hat{\sigma}})\, e^{\hat{\hat{\sigma}}} \hat{V} 
- \sum_{n>0} B_{n}\, \hat{\hat{\sigma}}^{n} \bar{V}^{(k)}_{R}
\label{eq:26}
\end{equation}
The above expression allows the construction of the $\bar{H}$  only from knowledge of the rest of the part of the similarity-transformed fluctuation potential. 
One can start with the guess $\bar{V}^{(0)}_{R} = \hat{V}_{R}$ and put into Eq. (\ref{eq:26}) to obtain
\begin{align}
\bar{V}^{\{1\}}_{R} 
= \hat{\hat{\sigma}}\hat{F} \nonumber \\
+ \left( 1 - \tfrac{1}{2}\hat{\hat{\sigma} }+ \tfrac{1}{12}\hat{\hat{\sigma}}^{2} + \cdots \right)
\left( 1 + \hat{\hat{\sigma} }+ \tfrac{1}{2}\hat{\hat{\sigma}}^{2} + \cdots \right) \hat{V} \nonumber \\
- \left( -\tfrac{1}{2}\hat{\hat{\sigma}} + \tfrac{1}{12}\hat{\hat{\sigma}}^{2} + \cdots \right) {\hat{V}}_{R} \nonumber \\
=\bigl( \hat{\hat{\sigma}}\hat{F} + \hat{V} + \hat{\hat{\sigma}}\hat{V} + \tfrac{1}{2}\hat{\hat{\sigma}}^{2}\hat{V} + \cdots \bigr)\nonumber \\
+ \bigl( -\tfrac{1}{2}\hat{\hat{\sigma}}\hat{V} - \tfrac{1}{2}\hat{\hat{\sigma}}^{2}\hat{V} - \tfrac{1}{4}\hat{\hat{\sigma}}^{3}\hat{V} - \cdots \bigr)\nonumber \\
+\left( \tfrac{1}{12}\hat{\hat{\sigma}}^{2}\hat{V} 
+ \tfrac{1}{12}\hat{\hat{\sigma}}^{3}\hat{V} 
+ \tfrac{1}{24}\hat{\hat{\sigma}}^{4}\hat{V} + \cdots \right)\nonumber \\
+ \left( \tfrac{1}{2}\hat{\hat{\sigma}}\hat{V}_{R} 
- \tfrac{1}{12}\hat{\hat{\sigma}}^{2}\hat{V}_{R} - \cdots \right) \nonumber \\
=\hat{\hat{\sigma}}\hat{F}+ \hat{V}
+ \hat{\hat{\sigma}}\hat{V}
- \tfrac{1}{2}\hat{\hat{\sigma}}\hat{V}
+ \tfrac{1}{2}\hat{\hat{\sigma}}\hat{V}_{R} \nonumber \\
+ \tfrac{1}{2}\hat{\hat{\sigma}}^{2}\hat{V}
- \tfrac{1}{2}\hat{\hat{\sigma}}^{2}\hat{V}
+ \tfrac{1}{12}\hat{\hat{\sigma}}^{2}\hat{V}
- \tfrac{1}{12}\hat{\hat{\sigma}}^{2}\hat{V}_{R}
+ \cdots \nonumber \\
= \hat{\hat{\sigma}}\hat{F} + \hat{V}
+ \tfrac{1}{2}\,\hat{\hat{\sigma}}\hat{V} + \tfrac{1}{2}\,\hat{\hat{\sigma}}\hat{V}_{R}
+ \tfrac{1}{12}\,\hat{\hat{\sigma}}^{2}\,\hat{V}_{\mathrm{N}} + \cdots
\label{eq:27}
\end{align}
Which can then be plugged into Eq. (\ref{eq:26}) to obtain $\bar{V}_{R}^{(2)}$ and so on.
Therefore, the total similarity-transformed Hamiltonian can be expressed as 
\begin{equation}
\bar{H} = \sum_{\mu} \bar{H}_{\mu}
\label{eq:28}
\end{equation}
where
\begin{equation}
\bar{H}_{0} = \hat{F} + \hat{V}
\label{eq:29}
\end{equation}
\begin{equation}
\bar{H}_{1} = [\hat{F}, \hat{\sigma}] 
+ \tfrac{1}{2}[\hat{V}, \hat{\sigma}] 
+ \tfrac{1}{2}[\hat{V}_{R}, \hat{\sigma}]
\label{eq:30}
\end{equation}
\begin{equation}
\bar{H}_{2} 
= \tfrac{1}{12}\bigl[[\hat{V}_{N}, \hat{\sigma}], \hat{\sigma}\bigr] 
+ \tfrac{1}{4}\bigl[[\hat{V}, \hat{\sigma}]_{R}, \hat{\sigma}\bigr] 
+ \tfrac{1}{4}\bigl[[\hat{V}_{R}, \hat{\sigma}]_{R}, \hat{\sigma}\bigr]
\label{eq:31}
\end{equation}
\begin{align}
\bar{H}_{3} =
\tfrac{1}{24}\bigl[[[\hat{V}_{N}, \hat{\sigma}], \hat{\sigma}]_{R}, \hat{\sigma}\bigr]
+ \tfrac{1}{8}\bigl[[[\hat{V}_{R}, \hat{\sigma}]_{R}, \hat{\sigma}]_{R}, \hat{\sigma}\bigr]\nonumber \\
+ \tfrac{1}{8}\bigl[[[\hat{V}, \hat{\sigma}]_{R}, \hat{\sigma}]_{R}, \hat{\sigma}\bigr]
- \tfrac{1}{24}\bigl[[[\hat{V}, \hat{\sigma}]_{R}, \hat{\sigma}], \hat{\sigma}\bigr]\nonumber \\
- \tfrac{1}{24}\bigl[[[\hat{V}_{R}, \hat{\sigma}]_{R}, \hat{\sigma}], \hat{\sigma}\bigr]
\label{eq:32}
\end{align}
\begin{align}
\bar{H}_{4} =\tfrac{1}{16}\bigl[[[[\hat{V}_{R}, \hat{\sigma}]_{R}, \hat{\sigma}]_{R}, \hat{\sigma}]_{R}, \hat{\sigma}\bigr] 
+ \tfrac{1}{16}\bigl[[[[\hat{V}, \hat{\sigma}]_{R}, \hat{\sigma}]_{R}, \hat{\sigma}]_{R}, \hat{\sigma}\bigr]\nonumber \\
+ \tfrac{1}{48}\bigl[[[[\hat{V}_{N}, \hat{\sigma}], \hat{\sigma}]_{R}, \hat{\sigma}]_{R}, \hat{\sigma}\bigr] 
- \tfrac{1}{48}\bigl[[[[\hat{V}, \hat{\sigma}]_{R}, \hat{\sigma}], \hat{\sigma}]_{R}, \hat{\sigma}\bigr] \nonumber \\
- \tfrac{1}{48}\bigl[[[[\hat{V}_{R}, \hat{\sigma}]_{R}, \hat{\sigma}], \hat{\sigma}]_{R}, \hat{\sigma}\bigr] 
- \tfrac{1}{144}\bigl[[[[\hat{V}_{N}, \hat{\sigma}], \hat{\sigma}]_{R}, \hat{\sigma}], \hat{\sigma}\bigr] \nonumber \\
- \tfrac{1}{48}\bigl[[[[\hat{V}, \hat{\sigma}]_{R}, \hat{\sigma}]_{R}, \hat{\sigma}], \hat{\sigma}\bigr] 
- \tfrac{1}{48}\bigl[[[[\hat{V}_{R}, \hat{\sigma}]_{R}, \hat{\sigma}]_{R}, \hat{\sigma}], \hat{\sigma}\bigr] \nonumber \\
 -\tfrac{1}{720}\bigl[[[[\hat{V}_{N}, \hat{\sigma}], \hat{\sigma}], \hat{\sigma}], \hat{\sigma}\bigr] 
\label{eq:33}
\end{align}
and so on.
Non-perturbative approximations to the UCC method can be derived by including different orders of $\sum_{\mu} \hat{H}_{\mu},\mu=1,2,3,\cdots$ in the energy and amplitude equations. Among the various approximations offered by this approach, the qUCC method provides the best compromise between computational cost and accuracy.\cite{phillips2025comparing}  The qUCC energy and amplitudes are given by
\begin{equation}
\langle \Phi_0 |\bar {H}_{0}+\bar {H}_{1}+\bar {H}_{2}+\bar {H}_{3}| \Phi_0 \rangle = E_{0}
\label{eq:34}
\end{equation}
\begin{equation}
\langle \Phi_\mu |\bar {H}_{0}+\bar {H}_{1}+\bar {H}_{2}| \Phi_0 \rangle =0
\label{eq:35}
\end{equation}
The qUCC method is generally employed within the singles-and-doubles approximation (qUCCSD).\cite{liu2021unitary} Extensions of the qUCCSD method to include noniterative triples corrections have also been developed.\cite{majee2025perturbativetriplescorrectionrelativistic} 
Alternatively, perturbative approximations to UCC can be derived by truncating the amplitudes in  Eq. (\ref{eq:25}) according to perturbation order, and it is generally denoted UCCn, where n denotes the order of perturbation.  It is important to note that the UCCn approach differs from Bartlett's UCC(n) framework\cite{bartlett1989alternative}, where the ground state energy is accurate through the order n in MP perturbation theory. In the present formulation, the amplitude equations are consistent through the n$^{\text{th}}$ perturbation order.
\begin{comment}In both truncation strategies, the expansion of the  $(\bar{H})$ involves Bernoulli numbers as the expansion coefficient\cite{liu2018unitary}. In the UCC3 method, the BCH expansion equation \ref{eq:10} is consistently expanded up to the third order in perturbation theory, while in the qUCCSD scheme, the expansion is carried out up to the second commutator as shown in \ref{eq:10}. The details of the truncation scheme via both the perturbative and commutator-based approaches can be found in the reference.\cite{liu2018unitary,liu2022quadratic,majee2024reduced}
It is important to note that our UCCn approach differs from Bartlett's UCC(n) framework\cite{bartlett1989alternative}, where the ground state energy is accurate through order in n in MP perturbation theory. In our formulation, the single and double amplitude equations are each consistent through the nth order, and starting at third order, they diverge from UCC(n) results due to the use of Bernoulli's expansion. This UCCn approach is the same as those references.\cite{liu2018unitary,liu2022quadratic,hodecker2020unitary,majee2024reduced} 
\end{comment}

\subsection{\label{app:subsec}First order Property Calculation using the UCC Expectation-Value Formalism}
The wave function for the \textit{I}$^{th}$ excited state in the UCC formalism is defined as
\begin{equation}
\lvert \Psi_{I} \rangle = e^{\hat{\sigma}} \hat{C}_{I} \, \lvert \Phi_{0} \rangle
\label{eq:36}
\end{equation}
Here, $\hat{C}_{I} $ is a CI-like excitation operator. For the ground state, $\hat{C}_{I}$  is an identity operator. The corresponding energy is defined as 
\begin{equation}
E_{I} =\langle \Phi_{0} \lvert\hat{C}^{\dagger}_{I} e^{-\hat{\sigma}} \hat{H} \, e^{\hat{\sigma}}\hat{C}_{I}\rvert \Phi_{0}\rangle
\label{eq:37}
\end{equation}

The first derivative of the energy with respect to an arbitrary perturbation ($\chi$) can be written as
\begin{align}
\frac{\partial E_{I}}{\partial \chi}=\langle \Phi_{0} \lvert\hat{C}^{\dagger}_{I} \frac{\partial e^{-\hat{\sigma}} }{\partial \chi} \hat{H} \, e^{\hat{\sigma}}\hat{C}_{I}\rvert \Phi_{0}\rangle  \nonumber \\+\langle \Phi_{0} \lvert\hat{C}^{\dagger}_{I} e^{-\hat{\sigma}} \frac{\partial\hat{H} }{\partial \chi} \, e^{\hat{\sigma}}\hat{C}_{I}\rvert \Phi_{0}\rangle +\langle \Phi_{0} \lvert\hat{C}^{\dagger}_{I} e^{-\hat{\sigma}} \hat{H} \, \frac{\partial e^{\hat{\sigma}}}{\partial \chi }\hat{C}_{I}\rvert \Phi_{0}\rangle 
\label{eq:38}
\end{align}

All possible Slater determinants can be partitioned into a principal space ($ \rvert \Phi_{P} \rangle$) and a complementary space ($\rvert  \Phi_{Q} \rangle $).
\begin{equation}
\rvert \Phi_{P} \rangle=e^{\hat{\sigma}}\rvert\Phi_{0} \rangle
\label{eq:39}
\end{equation}

such that all ($ \rvert \Phi_{P} \rangle$) determinants spans the basis of $\bar{H}_{\text{qUCC}}$.

In the FCI limit, the above expression can be rewritten (see the Supporting Information for details) as 

\begin{equation}
\frac{\partial E^{qUCC}_{I}}{\partial \chi}=\langle \Phi_{P} \lvert\hat{C}^{\dagger}_{I}  \frac{\partial\hat{H} }{\partial \chi} \, \hat{C}_{I}\rvert \Phi_{P}\rangle
\label{eq:40}
\end{equation}
However, for the truncated space, this expression is only an approximation. 
Accordingly, the first-order property corresponding to an arbitrary operator ($\hat{D}=\frac{\partial\hat{H} }{\partial \chi}$) in the qUCCSD expectation-value formalism can be written as 
\begin{small}
\begin{align}   
 d_{IJ}  = \nonumber \\
 \langle \Phi_{0} \lvert \hat{C}^{\dagger}_{I}  \left(  \hat{D} + [\hat{D}, \hat{\sigma}] 
+ \tfrac{1}{2!} [[\hat{D}, \hat{\sigma}], \hat{\sigma}] +  \tfrac{1}{3!} [[[\hat{D}, \hat{\sigma}], \hat{\sigma}],\hat{\sigma}] \right) \hat{C}_{J} \rvert \Phi_{0} \rangle
\label{eq:41}
\end{align}
\end{small}

When $I=J$, the expression yields the first-order property corresponding to the $I^{th}$ excited state, whereas $I\neq J$ corresponds to a transition property between the $I^{th}$ and $J^{th}$ excited states.
Consequently, the ground-state first-order property at the qUCCSD level can be defined as
\begin{align}
d_{0} =\langle \Phi_{0} \lvert \hat{D} + [\hat{D}, \hat{\sigma}] 
+ \tfrac{1}{2!} [[\hat{D}, \hat{\sigma}], \hat{\sigma}] + \tfrac{1}{3!} [[[\hat{D}, \hat{\sigma}], \hat{\sigma}],\hat{\sigma}] \rvert \Phi_{0}\rangle 
\label{eq:42}
\end{align}

The above described approach for the first-order property calculation is similar in spirit to the ADC intermediate state representation\cite{PhysRevA.43.4647,UCCprop}.
The programmable expressions for the UCC3 and qUCCSD first-order ground state property calculations are provided in the Supporting Information. The corresponding UCC3 expression can be derived by neglecting terms containing amplitude products beyond fourth order in perturbation theory in Eq.(\ref{eq:42}). To assess the performance of the new relativistic UCC3 and qUCCSD expectation-value approaches, we calculated the permanent dipole moment (PDM), magnetic hyperfine structure (HFS) constant, and electric field gradient (EFG) of small molecules containing heavy elements and compared the results with those obtained using the CCSD Z-vector method and with available experimental data.

\subsubsection{\label{app:subsec}Permanent Dipole Moment (PDM)}
The PDM operator  ($\vec{\mu_{d}}$) corresponding to a molecular system  can be expressed as the sum of the electronic contribution and the nuclear contribution as,
\begin{equation}
{\vec{\mu_{d}}} = -\sum_{i} e \vec{r}_i + \sum_{A} Z_A e \vec{r}_A
\label{eq:43}
\end{equation}
In the above equation, \textit{e} is the charge of the electron, $\vec{r_{i}}$ is the position vector of $i^{th}$ electron from the 
origin, $Z_{A}$ is the atomic number and $\vec{r}_A$ is the position vector of $A^{th}$ nucleus. Now using Eq.(\ref{eq:43}) in Eq.(\ref{eq:42}) one can get the expectation value for the PDM, as 
\begin{equation}
\mu_{d} = \langle \Phi_0 | e^{-\hat{\sigma}} {(-\sum_{i} e \vec{r}_i)} e^{\hat{\sigma}} | \Phi_0 \rangle +\sum_AZ_{A}e\vec{r}_A
\label{eq:44}
\end{equation}
The electronic contribution to $\mu_{d}$ in Eq.(\ref{eq:44}) depends on the choice of correlation method and basis set, whereas the nuclear part is independent of both the correlation method and the basis set; it remains constant for a particular molecule with a certain charge and at a specific geometry.
%\begin{equation}
%PDM = NC + EC_{DHF}+EC_{corr}
%\label{eq:22}
%\end{equation}

\subsubsection{\label{app:subsec}Hyperfine Structure (HFS) Constant }
The magnetic hyperfine splitting in atoms arises from the interaction between the electromagnetic field generated by the electrons and the magnetic dipole moment of the nucleus. Within the magnetic dipole approximation, the vector potential $\vec{A}$ at the position of the electron $i$ due to the nuclear magnetic moment $(\vec{\mu}_{k})$ arising from nucleus k,is given by    

\begin{equation}
\vec{A}_{i}=\frac{\vec{\mu}_{k}\times \vec{r}}{r^3},
\label{eq:45}
\end{equation}
where $r_{i}$ is the position vector of the electron relative to the nucleus. In a relativistic framework, the corresponding magnetic hyperfine interaction for all electrons $(n)$ can be written as 
\begin{equation}
H_{\text{HFS}} = \sum_{i}^{n} \vec{\alpha}_i \cdot \vec{A}_i.
\label{eq:46}
\end{equation}
where $\alpha_{i}$ denotes the Dirac matrices of the electron $i$. 
For a diatomic molecule, the magnetic hyperfine interaction can be separated into parallel $(A_{\parallel})$ and perpendicular $(A_{\perp})$ components of the magnetic HFS constant. Along the molecular axes, the $\textit{z}-$ component of the expectation value of the corresponding HFS operator yields the parallel constant ($A_{\parallel}$) while the components perpendicular to the molecular axes ($\textit{x}$ and $\textit{y}$) give the perpendicular constants ($A_{\perp}$), expressed as \cite{BrownCarrington2003, FroschFoley1952} 
\begin{equation}
A_{\parallel(\perp)} = \frac{\vec{\mu}_{k}}{I \Omega} \cdot \left\langle \Psi_{0}  \left | \sum_{i}^{n} \left( \frac{\vec{\alpha}_{i} \times \vec{r}_{i}}{r_{i}^{3}} \right)_{z(x/y)}\right| \Psi_0 \right \rangle
\label{eq:47}
\end{equation}
Where $I$ denotes the nuclear spin quantum number and $\Omega$ is the projection of total electronic angular momentum along the $\textit{z}$-axis of the diatomic molecule.

\subsubsection{\label{app:subsec} Electric Field Gradient (EFG)}
The EFG $(V_{ij})$ is a traceless symmetric second-rank tensor that describes the spatial variation of the electric field around a nucleus due to the distribution of surrounding charges. The EFG provides valuable information about the local electronic structure and symmetry of atoms or molecules, particularly in cases where the environment of the nucleus is asymmetric. 
Within the principal axis coordinate system, the expectation value of the $zz$ component of EFG $(V_{zz})$  at the position ($\vec{R}_{K}$) of nucleus \textit{K} is defined as

\begin{equation}
\label{eq:48}
\scalebox{0.85}{$
\begin{aligned}
\langle {V}_{zz}(\vec{R}_{K}) \rangle = 
& \frac{e}{4\pi\epsilon_{0}} 
\left\langle \Psi_{0} \left| 
\sum_{i}
\left[
\frac{3(r_{iz}-R_{Kz})^2}{|\vec{r}-\vec{R}_{K}|^5}
-\frac{1}{|\vec{r}-\vec{R}_{K}|^3}
\right]
\right| \Psi_{0} \right\rangle \\
& -\frac{e}{4\pi\epsilon_{0}} 
\sum_{L\neq K} Z_{L}
\left[
\frac{3(R_{Lz}-R_{Kz})^2}{|\vec{R}_{L}-\vec{R}_{K}|^5}
-\frac{1}{|\vec{R}_{L}-\vec{R}_{K}|^3}
\right]
\end{aligned}
$}
\end{equation}
The $\langle {V}_{zz}(\vec{R}_{K}) \rangle$ defined above can be written as the sum of two contributions for a molecular system
\begin{equation}
\label{eq:49}
\left\langle V_{zz}\!\left(\vec{R}_{k}\right)\!\right\rangle
= 
\left\langle q_{zz}\!\left(\vec{R}_{k}\right)\!\right\rangle 
+ \Omega^{\mathrm{nuc}}\!\left(\vec{R}_{k}\right)
\end{equation}

The first term in the above equation is the electronic contribution, named $\langle q_{zz}(\vec{R}_{k})\rangle$, which depends on the ground state wave function.
The second term, nuclear contribution $\Omega^{nuc}(\vec{R}_{K})$, which arises from the other nuclei in the molecule, depends only on the position of that remaining nucleus and is independent of the electronic wave function.

\section{Computational and implementation details}
The four-component UCC3 and qUCCSD expectation-value approaches for the calculation of first-order properties have been implemented in our in-house quantum chemistry software package, BAGH.\cite{duttaBAGHQuantumChemistry2025_a} 
The package is primarily developed in Python, with computationally intensive components written in Fortran and Cython. BAGH relies on external software packages for the generation of one- and two-electron integrals and is interfaced with PySCF\cite{sun2020recent,sun2018pyscf,sun2015libcint}, socutils,\cite{socutils} DIRAC,\cite{saue_2025_14833106} and GAMESS-US.\cite{barca2020recent} The DIRAC package is used to solve the DHF equations and to compute one- and two-electron integrals, along with one-electron property integrals required for the PDM and HFS calculations. For the EFG calculations, the DHF orbitals and the required integrals are generated using the PySCF package.\cite{sun2020recent,sun2018pyscf,sun2015libcint} \\
Dipole moment calculations were performed for a series of alkaline earth metal monofluoride (AF, where A = Mg, Ca, Sr, Ba), employing a quadruple-zeta basis set: Dyall.cv4z\cite{dyall2009relativistic,dyall2016relativistic} for the metal atoms and cc-pCVQZ\cite{dunning1989gaussian} for fluorine.
For HFS constant calculations, the aug-cc-pCV5Z\cite{dunning1989gaussian} basis set was used for hydrogen, aug-cc-pCVQZ for F, Be, and Mg, and Dyall.cv4z for Ca. 
EFG calculations were carried out using Dyall.cv2z\cite{dyall2011relativistic} basis set for all FX molecules[where, X= F, Cl, Br, I, At]. 
%Frozen core approximation has not been used for the EFG calculations. \\

\section{Result and Discussion}

\subsection{Permanent Dipole Moment}
Table \ref{table:1} summarizes the permanent dipole moments (PDMs) of the Group II monofluoride molecules (MgF, CaF, SrF, and BaF) obtained using the UCC3, qUCCSD, and Z-vector methods, along with available experimental data. All calculated values are reported in Debye. To ensure consistency with the available Z-vector results, our calculations were performed using the same molecular geometries, basis sets, and cutoff energies as those employed by Pal and co-workers.\cite{haldar2021molecular} 

\begin{figure*}[] 
\includegraphics[width=1.0\textwidth]{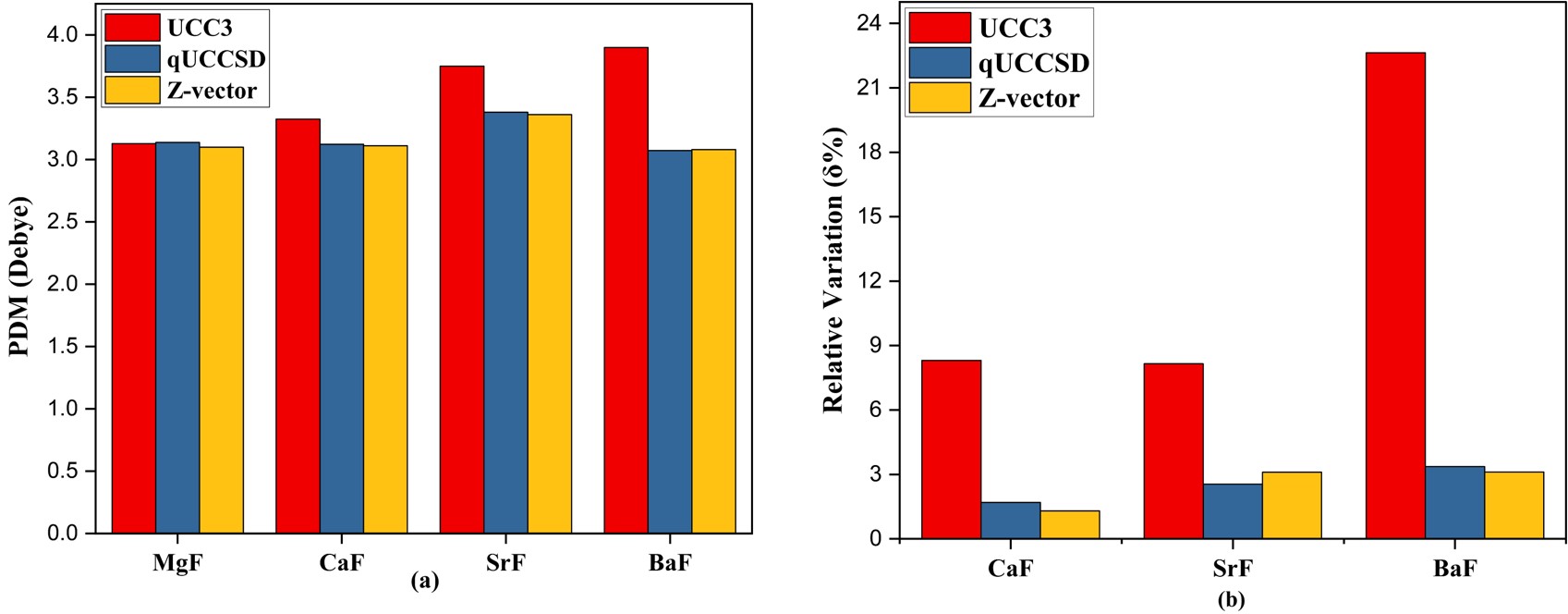} % Here is how to import EPS art
\caption{\text{(a)} Comparison of the total permanent dipole moments (PDMs) calculated using the four-component relativistic UCC3, qUCCSD, and Z-vector methods. \text{(b)} Relative deviations of the PDMs obtained using the UCC3, qUCCSD, and Z-vector methods from experimental values.}
\label{fig:my_label1}
\end{figure*}
Figure \ref{fig:my_label1}(a) compares the UCC3 and qUCCSD results with those obtained using the Z-vector method and shows that the qUCCSD results are in excellent agreement with the Z-vector approach. In contrast, the deviation obtained with the UCC3 method is significantly larger. This discrepancy arises because the qUCCSD method includes a more complete treatment of single excitations (up to quadratic terms in the amplitude equations), whereas these contributions are linear in the UCC3 amplitude equations. For MgF, the UCC3 and qUCCSD approaches agree remarkably well with each other; however, this agreement deteriorates as one moves down the group. In the case of CaF, the UCC3 method shows a deviation of 0.22 Debye from the Z-vector result, whereas the corresponding deviation for the qUCCSD method is only 0.01 Debye. Similarly, for SrF, the deviations from the CCSD Z-vector method are 0.39 Debye and 0.03 Debye for the UCC3 and qUCCSD, respectively. For BaF, the UCC3 method shows the largest deviation, amounting to 0.82 Debye, whereas the qUCCSD method maintains high accuracy with a deviation of only 0.01 Debye. These findings demonstrate that the qUCCSD method yields results in excellent agreement with those obtained using the Z-vector approach.

We have compared our results with available experimental measurements for CaF, SrF, and BaF. No experimental measurements are available for MgF. To assess the accuracy of our results, deviations from the experimental values were expressed as percentage errors, denoted by $\delta\%$ as
\begin{equation}
\delta \% = \left| \frac{\text{Expt.} - \text{Theory}}{\text{Expt.}} \right| \times 100 .
\label{eq:50}
\end{equation}
Figure \ref{fig:my_label1}(b) compares the percentage deviations ($\delta\%$) of the UCC3, qUCCSD, and Z-vector results from the experimental values. It can be seen that the UCC3 method displays larger deviations from the experimental values, whereas the qUCCSD and Z-vector methods yield more accurate results and are generally very close to each other. For CaF, the UCC3 method shows the smallest deviation from experiment, amounting to 8.15 $\%$.
 For SrF, the qUCCSD and Z-vector methods yield much lower deviations of 1.70$\%$ and 1.30$\%$, respectively. The UCC3 method shows the highest deviation for BaF, amounting to 22.63$\%$. For BaF, the qUCCSD and Z-vector results show very good agreement with experiment, with markedly improved deviations of 3.36$\%$, and 3.11$\%$,  respectively.
\begin{table*}[ht]
\caption{Comparison of total permanent dipole moments (PDMs, in Debye) obtained using the four-component relativistic UCC3 and qUCCSD methods with the CZ-vector method and available experimental data, along with the percentage deviations $(\delta\%)$ of each method from experiment.}
\begin{ruledtabular}
\begin{tabular}{cccccccc}
\multirow{2}{*}{Molecule} & \multirow{2}{*}{UCC3} & \multirow{2}{*}{qUCCSD} & \multirow{2}{*}{Z-vector\cite{haldar2021molecular}} & \multirow{2}{*}{Expt.} & \multicolumn{3}{c}{$\delta\%$}  \\
\cline{6-8}
      & & & &    & UCC3 &  qUCCSD  &  Z-vector \\ \hline
MgF              &-3.13 &  -3.14   & -3.10  & --&-- & -- &-- \\
CaF              &-3.33 &  -3.12   & -3.11  & 3.07\cite{dipolemomentofCaF}    & 8.31  & 1.70 & 1.30 \\
SrF              &-3.75 &  -3.39   & -3.36  & 3.4676\cite{ernst1985electricdipole} & 8.15  & 2.54 & 3.10  \\
BaF              &-3.90 &  -3.07   & -3.08  & 3.179\cite{ernst1986hyperfinedipole} & 22.63 & 3.36 & 3.11  \\
\end{tabular}
\end{ruledtabular}
\label{table:1}
\end{table*}

\subsection{Hyperfine Structure Constant}
Hyperfine structure (HFS) constants are sensitive to the quality of the wave function in the nuclear region and therefore provide a stringent test of the wave function obtained using a given method. Accordingly, we computed the HFS constants for a series of diatomic molecules, including BeH, MgH, CaH, and CaF, using both the UCC3 and qUCCSD methods.The computed parallel ($A_{\parallel}$) and perpendicular ($A_{\perp}$) components of the HFS constants are presented in Tables \ref{table:2} and \ref{table:3}, respectively. To ensure consistency, our results are compared with those obtained using the Z-vector method, employing identical basis sets, molecular geometries, and cutoff values as described in Ref.\cite{sasmal2016calculation}. 
The data presented in Tables \ref{table:2} and \ref{table:3} reveal that the qUCCSD results show better agreement with both Z-vector and the experimental values than the UCC3 results, which exhibit larger deviations from experiment. The maximum and minimum deviations of UCC3  and qUCCSD for $A_{\parallel}$ values from the Z-vector results occur for different systems. For UCC3, the largest deviation (92.87 MHz) from the Z-vector result occurs in $^1$H in MgH, while the smallest (0.82 MHz) is found for $^{25}$Mg in MgF. In contrast, the qUCCSD method shows a maximum deviation of 6.61 MHz for $^{19}$F for MgF and a minimum deviation of 0.89 MHz for $^{1}$H in CaH.
\begin{figure*}[] 
\includegraphics[width=1.0\textwidth]{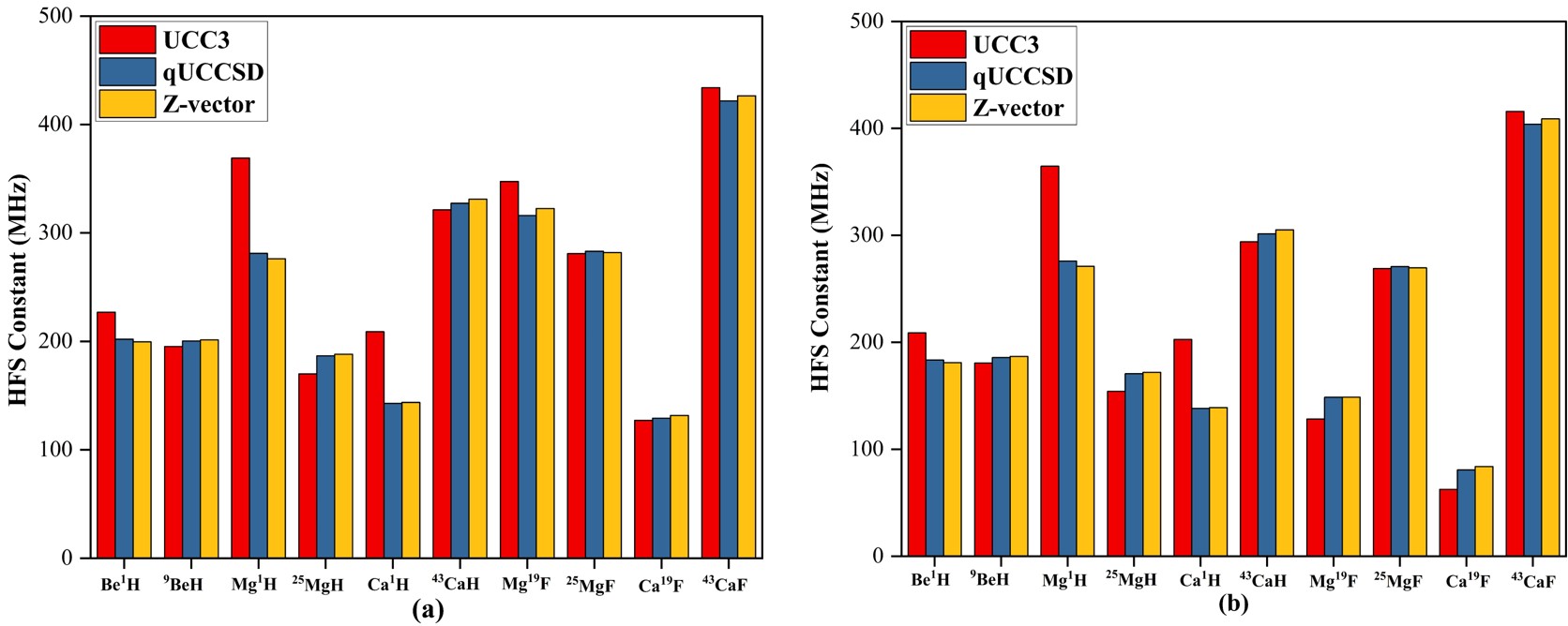} % Here is how to import EPS art
\caption{Comparison of the (a) $A_{\parallel}$ and (b) $A_{\perp}$ hyperfine-structure constants obtained using UCC3, qUCCSD and the Z-vector methods.}
\label{fig:my_label2}
\end{figure*}

\begin{figure*}[] 
\includegraphics[width=1.0\textwidth]{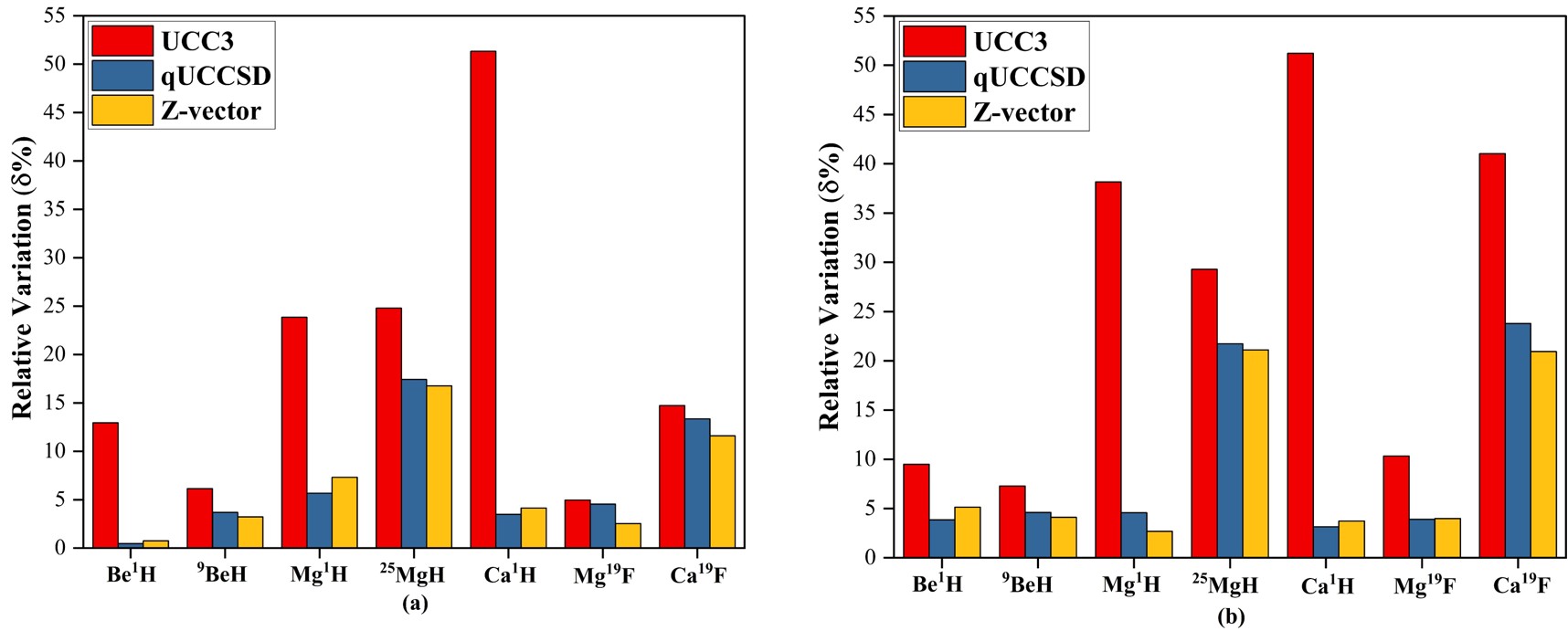} % Here is how to import EPS art
\caption{Relative deviations of the (a) $A_{\parallel}$ and (b) $A_{\perp}$ hyperfine-structure constants obtained using the UCC3, qUCCSD, and the Z-vector method with respect to the experimental values.}
\label{fig:my_label3}
\end{figure*}
In the case of $A_{\perp}$, the maximum deviation of 93.64 MHz from the Z-vector method is obtained in UCC3 again for $^1$H in MgH. The minimum deviation obtained with the UCC3 method (0.63 MHz) is observed for $^{25}$Mg in MgF. For the qUCCSD method, the maximum deviation from the Z-vector result is 4.98 MHz, observed for $^{25}$Mg in MgF. Figures \ref{fig:my_label2}(a) and \ref{fig:my_label2}(b) present comparisons of the $A_{\parallel}$ and $A_{\perp}$ values, respectively, obtained using the UCC3, qUCCSD, and Z-vector methods. The results clearly demonstrate that the qUCCSD method consistently yields better agreement with the Z-vector results than the UCC3 method.
Our calculated results were also compared with available experimental data. Figures \ref{fig:my_label3}(a) and \ref{fig:my_label3}(b) show the percentage deviation ($\delta\%$) of the $A_{\parallel}$ and $A_{\perp}$ values, respectively, obtained using UCC3, qUCCSD, and Z-vector methods. For the $A_{\parallel}$ component, the largest $\delta\%$ for the UCC3 method occurs for $^{1}$H in CaH, corresponding to a relative deviation of 51.34$\%$ from experiment. The higher relative deviations observed in the qUCCSD and Z-vector methods are 17.42 $\%$ and 16.77 $\%$, respectively, both occurring for the $^{25}$ Mg nucleus in MgH. 
In the case of the $A_{\perp}$ component, the largest deviation is observed for the $^1$H nucleus in CaH, amounting to 51.22 $\%$ for the UCC3 method. The qUCCSD approach shows its highest deviation of 23.78 $\%$, for the $^{19}$F nucleus in CaF, which is close to the corresponding Z-vector value of 20.94 $\%$. From the overall trends observed in the table, it is evident that the qUCCSD method demonstrates very good agreement with experimental values for both $A_{\parallel}$ and  $A_{\perp}$ components of the hyperfine coupling constant, showing close agreement with the corresponding Z-vector results. In contrast, the UCC3 method exhibits significant deviations from experimental results in several cases.

\begin{table*}[ht]
\caption{Comparison of hyperfine coupling constants ($ A_\parallel$,in MHz) calculated using different methods.}
\begin{ruledtabular}
\begin{tabular}{ccccccccc}
\multirow{2}{*}{Molecule} & \multirow{2}{*}{Atom} & \multirow{2}{*}{UCC3}& \multirow{2}{*}{qUCCSD} & \multirow{2}{*}{Z-vector\cite{sasmal2016calculation}} & \multirow{2}{*}{Expt.} & \multicolumn{3}{c}{$\delta\%$} \\
\cline{7-9}
          & & & & & &  UCC3  & qUCCSD  & Z-vector\\ \hline
BeH & $^1$H        &  227.02 &  201.94  &  199.5 & 201\cite{knight1972hyperfine}  & 12.94 & 0.47 & 0.75 \\
    & $^9$Be       & -195.21 & -200.35  & -201.3 & -208\cite{knight1972hyperfine} &  6.15 & 3.68 & 3.22 \\
 
MgH & $^1$H        &  369.07 &  281.12  &  276.2 & 298\cite{knight1971hyperfine1}  & 23.85  & 5.66  & 7.31 \\
    & $^{25}$Mg    & -169.97 & -186.64  & -188.1 & -226\cite{knight1971hyperfine1} & 24.79  & 17.42 & 16.77 \\

CaH & $^1$H        &  208.84 &  142.81  &  143.7 & 138\cite{knight1971hyperfine1}  & 51.34  & 3.49  & 4.31 \\
    & $^{43}$Ca    & -321.28 & -327.40  & -331.2 & -- & -- & -- & -- \\

MgF & $^{19}$F     &  347.41 &  315.99  &  322.6 & 331\cite{knight1971hyperfine2}   & 4.96   & 4.53  & 2.54 \\
    & $^{25}$Mg    & -280.98 & -283.02  & -281.8 & -- & -- & -- & -- \\

CaF & $^{19}$F     &  127.05 & 129.11   &  131.7 & 149\cite{knight1971hyperfine2}   & 14.73  & 13.35 & 11.61 \\
    & $^{43}$Ca    & -433.92 & -421.60  & -426.4 & -- & -- & -- & -- \\
\end{tabular}
\end{ruledtabular}
\label{table:2}
\end{table*}

\begin{table*}[ht!]
\caption{Comparison of hyperfine coupling constants ($A_\perp$,in MHz) calculated using different methods.}
\begin{ruledtabular}
\begin{tabular}{ccccccccc}
\multirow{2}{*}{Molecule} & \multirow{2}{*}{Atom} & \multirow{2}{*}{UCC3} & \multirow{2}{*}{qUCCSD}& \multirow{2}{*}{Z-vector\cite{sasmal2016calculation}} & \multirow{2}{*}{Expt.} & \multicolumn{3}{c}{$\delta\%$} \\
\cline{7-9}
               & & & & & &  UCC3   & qUCCSD  & Z-vector\\ \hline
BeH  & $^1$H             &   208.90   &  183.43  &  181.0  &  190.8\cite{knight1972hyperfine}  & 9.49 & 3.86  & 5.14 \\
     & $^9$Be            &  -180.61   & -185.83  & -186.8  & -194.8\cite{knight1972hyperfine}  & 7.28  & 4.60  & 4.11 \\

MgH  & $^1$H             &   364.74   &  276.08  &  271.1  &  264\cite{knight1971hyperfine1}   & 38.16 & 4.58  & 2.69 \\
     & $^{25}$Mg         &  -154.15   & -170.61  & -172.0  & -218.0\cite{knight1971hyperfine1} & 29.29 & 21.74 & 21.10 \\

CaH  & $^1$H             &   202.64   &  138.22  &  139    &  134\cite{knight1971hyperfine1}   & 51.22 & 3.15 & 3.73 \\
     & $^{43}$Ca         &  -293.97   & -301.46  & -305.2  &  --& -- & -- & -- \\

MgF  & $^{19}$F          &   128.23   &  148.59  &  148.7  &  143\cite{knight1971hyperfine2}    & 10.33 & 3.91 & 3.99 \\
     & $^{25}$Mg         &  -268.97   & -270.82  & -269.6  &  --& -- & -- & -- \\

CaF  & $^{19}$F          &   62.51    &  80.79   &  83.8   &  106\cite{knight1971hyperfine2}    & 41.03 & 23.78 & 20.94 \\
     & $^{43}$Ca         &  -415.89   & -403.98  & -408.9  & -- & -- & -- & -- \\
\end{tabular}
\end{ruledtabular}
\label{table:3}
\end{table*}

\subsection{Electric Field Gradient}

\begin{figure*}[ht!]
\includegraphics[width=1.0\textwidth]{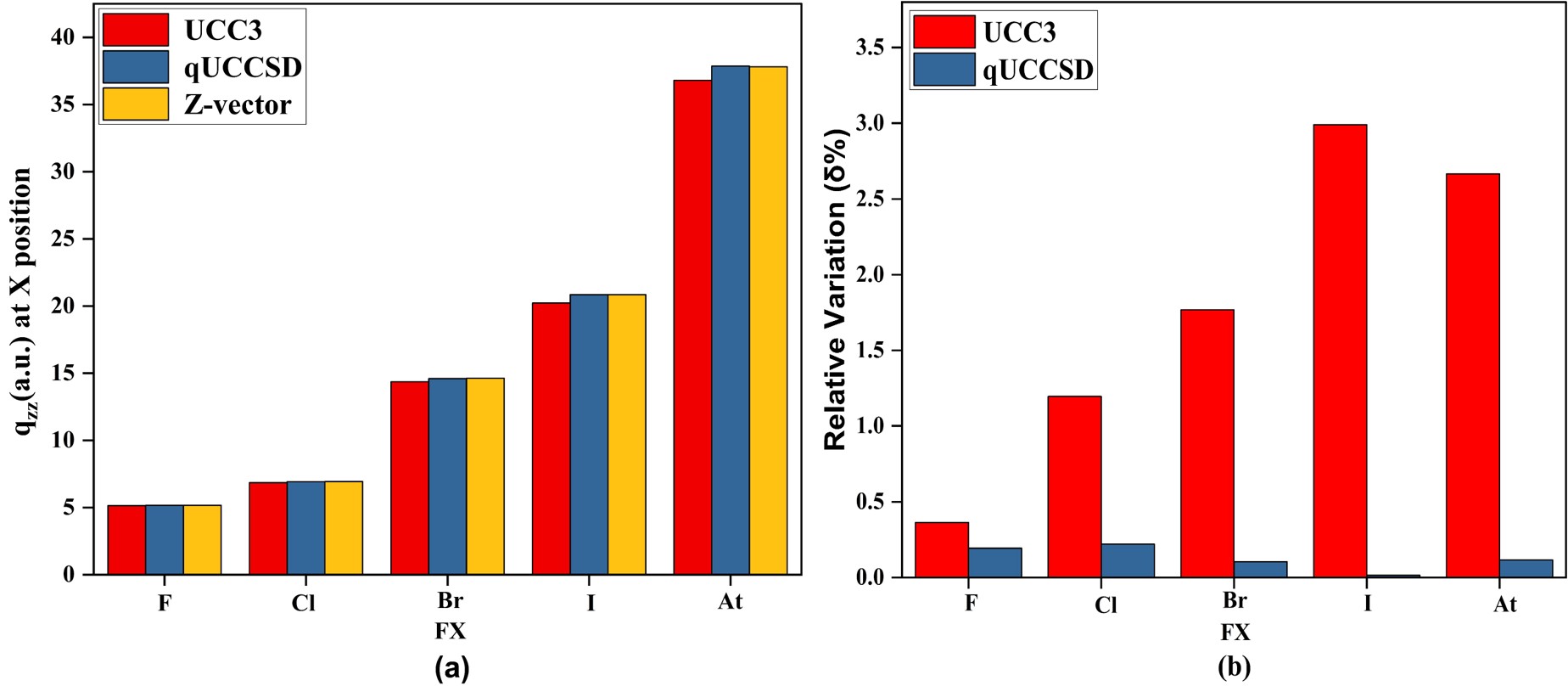} % Here is how to import EPS art
\caption{(a) Comparison of the $q_{zz}(\vec{R}_{X})$ values (in a.u.) for FX (X = F, Cl, Br, I, At) molecules obtained using the four-component UCC3, qUCCSD, and Z-vector methods. (b) Relative variation of $q_{zz}(\vec{R}_{X})$ values obtained using the UCC3 and qUCCSD methods with respect to the Z-vector results.}
\label{fig:my_label4}
\end{figure*}

To further assess the performance of the newly developed UCC methods, we calculated the electronic component ($q_{zz}$) of the electric field gradient ($V_{zz}(\vec{R}_{k})$), for a series of dihalogen molecules, FY (Y = F, Cl, Br, I, At), and compared them with the four-component CCSD results. The reference Z-vector values were taken from the work of Aucar \textit{et al.}.\cite{aucar2021relativistic} To maintain consistency, all calculations were carried out using the same Dyall.cv2z basis set and molecular geometries as employed in Ref. \onlinecite{aucar2021relativistic}. The molecular geometries for the series of dihalogen molecules are also provided in the Supporting Information. Figure \ref{fig:my_label4}(a) compares the UCC3 and qUCCSD values with the reference Z-vector results for the \textit{zz}-component (with the \textit{z}-axis chosen along the molecular bond) of the EFG values ($q_{zz}(\vec{R}_{k})$), for atoms in the FY series. The corresponding numerical values are listed in Table \ref{table:4}. It can be seen that the qUCCSD values closely follow the CCSD results throughout the data set, with the bars nearly overlapping in most cases. In contrast, the UCC3 method displays small but noticeable deviations, particularly for heavier atoms such as I and At. Table \ref{table:4} also reports the deviations($\delta\%$) from the reference values, which in this case correspond to the Z-vector results, as no experimental data are available. From Table \ref{table:4}, it is clear that the qUCCSD values show smaller deviations from the CCSD results than the UCC3 values, often by an order of magnitude. The largest deviation obtained with the UCC3 method is 1.008 a.u., observed for At in FAt. In contrast, the qUCCSD method shows a deviation of only 0.044 a.u. for the same system.

% Fig.~\ref{fig:epsart}%

\begin{table*}[ht!]
\caption{Comparison of the $q_{zz}(\vec{R}_{K})$ values (in a.u.) for FY (Y = F, Cl, Br, I, At) molecules obtained using the four-component UCC3 and qUCCSD methods with the corresponding Z-vector values, along with their relative deviations $(\delta\%)$.}  
\begin{ruledtabular}
\begin{tabular}{cccccccc}
\multirow{2}{*}{Molecule} & \multirow{2}{*}{Atom} & \multirow{2}{*}{UCC3} & \multirow{2}{*}{qUCCSD }& \multirow{2}{*}{Z-vector\cite{aucar2021relativistic}} & \multicolumn{2}{c}{$\delta\%$} \\
\cline{6-7}
& & &       &                  &  UCC3     & qUCCSD  \\ \hline 
F\textsubscript{2}    & F      &  5.145    &  5.174   &   5.164    & 0.364   & 0.194   \\
FCl                   & F      &  3.013    &  2.974   &   2.967    & 1.554   & 0.220   \\
                      & Cl     &  6.850    &  6.918   &   6.933    & 1.196   & 0.221   \\
FBr                   & F      &  1.872    &  1.773   &   1.768    & 5.892   & 0.280   \\
                      & Br     & 14.363    & 14.607   &  14.622    & 1.769   & 0.105   \\
IF                    & F      &  0.810    &  0.677   &   0.683    & 18.563  & 0.933   \\
                      & I      & 20.233    & 20.85    &  20.857    &  2.991  & 0.015   \\
FAt                   & F      & -0.251    & -0.531   &  -0.501    & 49.936  & 4.331   \\
                      & At     & 36.803    & 37.855   &  37.811    &  2.667  & 0.116   \\ 
\end{tabular}
\end{ruledtabular}
\label{table:4}
\end{table*}

\section{Conclusion}
We present the theory, implementation, and benchmarking of first-order property calculations using the relativistic unitary coupled-cluster expectation-value method. The Hermitian formulation within the qUCCSD method allows a simple definition of first-order properties using the expectation-value approach and does not require the calculation of an additional set of amplitudes, as required in the standard coupled cluster Z-vector method. We calculated the permanent dipole moment, hyperfine structure constant, and electric field gradient for a series of molecules containing heavy atoms. The values obtained using the qUCCSD expectation-value approach are in close agreement with the CCSD Z-vector results and are consistent with available experimental data. However, the perturbative UCC3 method does not exhibit consistent performance and often shows large errors. This behavior can presumably be attributed to the incomplete treatment of single excitations at the UCC3 level.
The present work demonstrates that the projection-based approximation to the UCC method provides an attractive option for the calculation of first-order properties, even on a classical computer. It would be interesting to extend the unitary coupled cluster method to second- and higher-order properties. Work in this direction is currently in progress. 
\section*{Supplementary Material}
\label{sec6}
The Supplementary Material contains the programmable expressions for the ground state first-order property in the UCC3 and qUCCSD methods, the derivation of the equivalence of the energy derivative and expectation value approach in UCC at the full CI limit, the molecular geometries for the series of dihalogen molecules, and the nuclear spins and magnetic moments of isotopes used in the calculations of the hyperfine structure constant.
\begin{acknowledgments}
AKD, SC, and KM acknowledge the support from IIT Bombay, ANRF(CRG/2023/002558), and ISRO  for financial support. The authors also acknowledge the IIT Bombay supercomputing facility and C-DAC (Param Smriti and Param Bramha) for providing computational resources. SC acknowledges the Prime Minister's Research Fellowship (PMRF), and KM acknowledges CSIR-HRDG for the research fellowship. AKD acknowledges the research fellowship funded by the EU NextGenerationEU through the Recovery and Resilience Plan for Slovakia under project No. 09I03-03-V04-00117. The authors acknowledge Juan J. Aucar (Institute for Modelling and Innovative Technology, IMIT (CONICET-UNNE), Argentina) for his valuable feedback and discussions on EFG calculations during the preparation of this manuscript. 
\end{acknowledgments}
\newpage
\renewcommand{\refname}{References} % For article class
\bibliographystyle{aipnum4-1}
%\bibliography{aipsamp}% Produces the bibliography via BibTeX.

\end{document}